\documentclass{iopart}
\usepackage{graphicx}
\usepackage{iopams}
\usepackage[square,comma,sort&compress]{natbib}

\begin{document}

\title[]{Fundamental relation between longitudinal and transverse conductivities in the quantum Hall system}

\author{Akira Endo$^1$, Naomichi Hatano$^2$, Hiroaki Nakamura$^3$ and Ry\=oen Shirasaki$^4$}

\address{$^1$ Institute for Solid State Physics, University of Tokyo, Kashiwanoha, Kashiwa, Chiba 277-8581, Japan\\
$^2$ Institute of Industrial Science, University of Tokyo, Komaba, Meguro, Tokyo 153-8505, Japan\\
$^3$ Department of Simulation Science, National Institute for Fusion Science, Oroshi-cho, Toki, Gifu 509-5292, Japan\\
$^4$ Department of Physics, Yokohama National University, Tokiwadai, Hodogaya-ku, Yokohama, Kanagawa 240-8501, Japan
}

\ead{akrendo@issp.u-tokyo.ac.jp}
\begin{abstract}
We investigate the relation between the diagonal ($\sigma_{xx}$) and off-diagonal ($\sigma_{xy}$) components of the conductivity tensor in the quantum Hall system. We calculate the conductivity components for a short-range impurity potential using the linear response theory, employing an approximation that simply replaces the self-energy by a constant value $-i \hbar /(2 \tau)$ with $\tau$ the scattering time. The approximation is equivalent to assuming that the broadening of a Landau level due to disorder is represented by a Lorentzian with the width $\Gamma = \hbar /(2 \tau)$. Analytic formulas are obtained for both $\sigma_{xx}$ and $\sigma_{xy}$ within the framework of this simple approximation at low temperatures. By examining the leading terms in $\sigma_{xx}$ and $\sigma_{xy}$, we find a proportional relation between $\mathrm{d}\sigma_{xy}/\mathrm{d}B$ and $B \sigma_{xx}^2$. The relation, after slight modification to account for the long-range nature of the impurity potential, is shown to be in quantitative agreement with experimental results obtained in the GaAs/AlGaAs two-dimensional electron system at the low magnetic-field regime where spin splitting is negligibly small.
\end{abstract}

\maketitle

\section{Introduction\label{Intr}}
The experimental finding by Chang and Tsui \cite{Chang85} of the striking similarity between the longitudinal resistivity $\rho_{xx}$ and the derivative of the Hall resistivity with respect to the electron density $n_e$, $\mathrm{d}\rho_{xy}/\mathrm{d}n_e$, in the quantum Hall regime has attracted considerable interest and has since been a subject of a number of experimental \cite{Rotger89,Morawicz90,Stormer92,Coleridge94,Allerman95,Tieke97,Pan05} and theoretical \cite{Vagner88,Simon94,Ilan06} studies. Using a low-carrier-density ($n_e$ $\leq$1$\times$10$^{-15}$ m$^{-2}$) high-mobility ($\mu$ $\geq $ 300 m$^2$V$^{-1}$s$^{-1}$) two-dimensional electron system (2DES) in GaAs/AlGaAs, Stormer \textit{et al.} \cite{Stormer92} showed that all features in $\rho_{xx}$ (including overshooting flanks around quantum Hall states) are faithfully reproduced by the derivative of $\rho_{xy}$ with respect to the magnetic field $B$ in the form
\begin{equation}
\label{chang}
B \frac{\mathrm{d} \rho_{xy}}{\mathrm{d} B} \simeq \beta \rho_{xx},
\end{equation}
where $\beta$ is a sample-dependent constant value (typically between 20 and 40). Note, as pointed out in Ref.\ \cite{Chang85}, that the differentiation by $B$ and that by $n_e$ are basically equivalent to each other, $-B (\mathrm{d}/\mathrm{d}B) = n_e (\mathrm{d}/\mathrm{d}n_e)$, if the relevant variable in the problem is the filling factor $\nu = n_eh/eB$ and not $n_e$ or $B$ separately. 

%%%%%%%%%%%%%%%%%%%%%%%%%%%%%%%%%%%%
%The value of $\beta$ is later explained by Coleridge \textit{et al.} \cite{Coleridge} in terms of the ratio between the momentum relaxation time $\tau$ and the quantum scattering time $\tau_q$, $\beta$=$2\tau /\tau_q$; in GaAs/AlGaAs 2DES, $\tau_q$ is known to be much smaller than $\tau$ owing to the long-range nature of the impurity scattering. Theoretically, Vagner {\it et al.} \cite{Vagner} applied a scaling approach to the problem. Their analysis leads to $n_e$ $\propto$ $B^0$ and $\tau$ $\propto$ $B^{-1}$ for Eq.\ (\ref{chang}) to be valid; the exponents are consistent with the long-range impurity scattering. 
%%%%%%%%%%%%%%%%%%%%%%%%%%%%%%%%%%%%%

The origin of the intriguing empirical relation Eq.\ (\ref{chang}) remains largely enigmatic. A possible explanation is given by Simon and Halperin \cite{Simon94}, who ascribed the relation to the microscopic inhomogeneity in the electron density $n_e$ inevitably present in real 2DES samples. Noting that the macroscopic value of $\rho_{xx}$ measured in experiments is mainly determined by the fluctuation in the local Hall resistivity $\rho_{xy}(\vec{r})$ resulting from the inhomogeneity in $n_e$ rather than by the local longitudinal resistivity $\rho_{xx}(\vec{r})$, their theory leads to Eq.\ (\ref{chang}) for not too low temperatures if disorders are taken into consideration on multiple length scales. In a recent experiment by Pan \textit{et al.} \cite{Pan05} using an ultrahigh mobility ($\mu$ = 3100 m$^2$V$^{-1}$s$^{-1}$) 2DES at an extremely low temperature ($\sim$ 6 mK), experimentally measured value of $\rho_{xx}$ was interpreted \cite{Pan05,Ilan06} as essentially reflecting the difference in $\rho_{xy}(\vec{r})$ between the voltage probes placed under slightly different ($\sim$0.5 \%) electron density $n_e$, and accordingly as virtually irrelevant to the local resistivity $\rho_{xx}(\vec{r})$. Note, however, that the van der Pauw geometry used in their study is not necessarily an ideal setup for the measurement of the resistivity. 

In the present paper, we explicitly calculate the diagonal ($\sigma_{xx}$) and the off-diagonal ($\sigma_{xy}$) components of the conductivity tensor in the quantum Hall system by employing the linear response theory. Although there already exist a number of sophisticated theories devoted to the calculation of $\sigma_{xx}$ and $\sigma_{xy}$ in the quantum Hall system (see, e.g., Refs.\ \cite{AndoOC74,Gerhardts75,Ishihara86,AndoRev82}), they have not been applied, to the knowledge of the present authors, to the interpretation of the relation between the two components of the conductivity tensor exemplified by Eq.\ (\ref{chang}). We take the effect of disorder into account by simply assuming the Lorentzian broadening of the Landau levels with the width $\Gamma$ independent of $B$; this can readily be done by substituting a constant value  $-i \Gamma$ for the self-energy in the Green's function. Although this appears to be somewhat an oversimplified approximation, the Lorentzian with the $B$-independent width is suggested by a number of experiments to be a function that describes quite well the broadening of Landau levels due to disorder \cite{Ashoori92,Potts96,Zhu03,Dial07}. By contrast, the well-known self-consistent Born approximation \cite{AndoOC74} yields a semi-elliptical broadening, which is by far a less accurate representation of the experimentally observed Landau levels. A great advantage of the simple approximation employed in the present study is that it allows us to deduce analytic formulas for both $\sigma_{xx}$ and $\sigma_{xy}$ for low enough temperatures $k_\mathrm{B} T$ $\ll$ $\varepsilon_\mathrm{F}$ with $\varepsilon _\mathrm{F}$ the Fermi energy. The analytic formulas, in turn, provide us with a transparent way to examine the underlying relation between the two components. By picking out the most significant terms at high magnetic fields in the formulas, we find the relation
\begin{equation}
\frac{\mathrm{d}\sigma_{xy}}{\mathrm{d}B} \simeq \lambda B \sigma_{xx}^2, 
\label{FR1}
\end{equation}
with the coefficient $\lambda$ determined by scattering parameters and $\varepsilon_\mathrm{F}$ [see Eq.\ (\ref{eq22}) below for details]. The relation is analogous to Eq.\ (\ref{chang}) but with a notable difference that $\sigma_{xx}$ enters the equation in squared form. Note that Eq.\ (\ref{chang}) can be rewritten as
\begin{equation}
B\frac{\mathrm{d}\sigma_{xy}}{\mathrm{d}B} \simeq \beta \sigma_{xx}
\label{chang2}
\end{equation}
by using the approximate relations for not too-small magnetic fields, $\rho_{xy}$ $\approx$ $1/\sigma_{xy}$ and $\rho_{xx}$ $\approx$ $\sigma_{xx}/\sigma_{xy}^2$. In contrast to the previous study \cite{Simon94}, we have not introduced inhomogeneity in $n_e$ in our calculation.

The relation between $\sigma_{xx}$ and $\sigma_{xy}$ found in the present study is compared with experimental results obtained in a GaAs/AlGaAs 2DES using the Hall-bar geometry, a geometry well-suited to the measurement of the resistivity. Care should be taken in the comparison, since our theoretical calculation is based on the short-range impurity potential, while the dominant scattering in a GaAs/AlGaAs 2DES is known to be of long-ranged. We find that Eq.\ (\ref{fundrelLRn}) below obtained by modifying Eq.\ (\ref{FR1}) to accommodate the long-range potential describes the experimental results remarkably well for the low magnetic field range where the spin splitting, the localization, the formation of edge states, and the electron-electron interaction can be neglected.

The paper is organized as follows. In Sec.\ \ref{ImpSc}, we introduce the Green's function to be employed in the later calculations. Components of the conductivity tensor are calculated in Sec.\ \ref{CondTen}, which are shown in \ref{WeakMag} to approach the semiclassical formulas asymptotically for $B \rightarrow 0$. The relation between $\sigma_{xx}$ and $\sigma_{xy}$ is examined in Sec.\ \ref{Relation}, and is compared with experimental results in Sec.\ \ref{CompExp} after modification to account for the long-range nature of the impurity potential. The validity of our approximation and the magnetic-field range for our approximation to be accurate are discussed in Sec.\ \ref{Disc}, followed by concluding remarks in Sec.\ \ref{Conc}.

\section{Impurity scattering in the quantum Hall system\label{ImpSc}}
We consider a 2DES in a magnetic field perpendicular to the 2D plane. The Hamiltonian of the system is given by
\begin{eqnarray}
\label{H}
 H_{\rm QH} & = & H_0 + V_{\rm imp},
 \\
\label{H0}
 H_0 & = & \frac{1}{2m^*} (\vec{p} + e \vec{A})^2,
\end{eqnarray}
where $\vec{p}$ denotes the momentum operator, $-e$ is the charge of an electron, $\vec{A}$ is the vector potential of the magnetic field $(0,0, B)$ and $V_{\rm imp}$ represents the impurity potential. We neglect spins for simplicity. The term $H_0$ in the Hamiltonian gives the Landau levels. The eigenfunction of $H_0$ in the Landau gauge is given by
\begin{equation}
\phi_{kN}(x,y) = \frac{1}{\sqrt{L}}e^{ikx} \chi_N (y-y_k), 
\label{eqwf}
\end{equation}
where $L$ is the length of the system, $\chi_N$ denotes the eigenfunction of the harmonic oscillator in the $N$th Landau level whose energy is given by $E_N=\hbar\omega_c (N+1/2)$ with $\omega_c=e |B|/m^*$ the cyclotron frequency, and $y_k=-k \ell ^2$ is the guiding center with $\ell = \sqrt{\hbar/e|B|}$ the magnetic length.

We consider a short-range potential of the form 
\begin{equation}
\label{eq5a}
 V_{\rm imp} (\vec{r}) = \sum_i V_i \delta (\vec{r}-\vec{r}_i).
\end{equation}
%%%%%%%%%%%%%%%%%%%%%%%%%%%%%%%%%
Owing to the impurity potential, Landau levels acquire width, which are otherwise delta functions placed at $\varepsilon=E_N$ ($N$ = 0, 1, 2,...). The resulting density of states (DOS), or the line shape of the impurity-broadened Landau levels, has been calculated for various types of impurity potential. For a white-noise potential (impurities with constant strength $V_i$ distributed at random positions $\vec{r_i}$), the broadening was shown to be well described by a Gaussian line shape \cite{Wegner83,Brezin84,Benedict87}. Calculations were also done assuming a distribution $P(V_i)$ in the strength of the impurity scattering $V_i$ \cite{Affleck84,Brezin84,Benedict87,Benedict85,Benedict86,Ando85,Raikh93}. Brezin \textit{et al.} \cite{Brezin84} and Benedict \textit{et al.} \cite{Benedict86} showed that a Lorentzian distribution of $P(V_i)$ results in DOS described by a Lorentzian line shape. Lorentzian broadening of the Landau levels is consistent with experiments on the tunneling into a 2DES \cite{Ashoori92,Dial07} or measurement of the magnetization in a 2DES \cite{Potts96,Zhu03}.

In the present paper, we start by assuming the Lorentzian DOS
\begin{equation}
\label{eq2a}
D (\varepsilon)  =  \frac{1}{2 \pi \ell^2}\sum_{N=0}^{\infty} \frac{1}{\pi} \frac{\Gamma}{(\varepsilon-E_N)^2+\Gamma^2} .
\end{equation}
As will be shown, this simple approximation allows us to deduce analytic formulas of the conductivity tensor, which proves to be essential for the later analysis of the relation between the components of the conductivity tensor. 

The simple DOS Eq.\ (\ref{eq2a}) implies analogous simplicity in the electron Green's function. For sufficiently short-ranged impurity potential, the Green's function can be written in the diagonalized form as
\begin{equation}
\label{eq3}
 G_N(\varepsilon) \delta_{N,N'}\delta_{k,k'} = \left< N,k \left| \frac{1}{\varepsilon-H_{\rm QH}} \right|N',k'\right>=\frac{\delta_{N,N'}\delta_{k,k'}}{\varepsilon-E_N - \Sigma_N (\varepsilon) },
\end{equation}
where $|N,k\rangle$ represents the eigenstate of the unperturbed Hamiltonian given by Eq.~(\ref{eqwf}), and $ \Sigma_N (\varepsilon)$ denotes the self-energy resulting from $V_{\rm imp}$.  The DOS is related to the imaginary part of the electron Green's function (\ref{eq3}) by
\begin{equation}
\label{eq1}
D(\varepsilon)  =  - \frac{ 1 }{{2\pi \ell^2}}\sum_N 
\rho_N (\varepsilon) ,
\end{equation}
with $\rho_N (\varepsilon)$ introduced as
\begin{equation}
\label{eq2}
\rho_N (\varepsilon)  =  \frac{1}{\pi} {\rm Im} G_N(\varepsilon + i0) .
\end{equation}
It is easy to see that Eq.\ (\ref{eq1}) reproduces Eq.\ (\ref{eq2a}) if the self-energy $\Sigma_N (\varepsilon)$ in Eq.\ (\ref{eq3}) is replaced by a constant value $- i \Gamma = - i \hbar / (2\tau)$, yielding
\begin{equation}
\label{eq4a}
G_N(\varepsilon + i 0) = \frac{1}{\varepsilon-E_N+i \Gamma}.
\end{equation}
We exploit the simple Green's function Eq.\ (\ref{eq4a}) in the following calculations.

\section{Conductivity tensor\label{CondTen}}

\begin{figure}[tb]
\begin{center}
\includegraphics[width=0.3\textwidth]{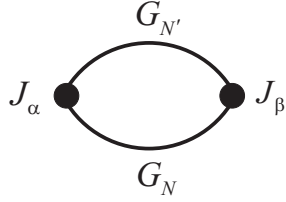}
\end{center}
\caption{\label{Fig-1}The diagram for the current-current correlation function. }
\end{figure}

We introduce the particle-current operator $\vec{j}$
%, using the velocity operator of the particle $\vec{v}$, 
of the form
% and the charge of the carrier $e$ as
%
\begin{equation}
%\vec{j} = \vec{v}=\frac{1}{m^*}(\vec{p}-e\vec{A}).
\vec{j} =\frac{1}{m^*}(\vec{p}+e\vec{A}).
\label{i0}
\end{equation}
The conductivity tensor $\sigma_{\alpha \beta}$ (with $\alpha$ and $\beta$ representing either of $x$ or $y$) of the 2DES is given by the Kubo formula
\begin{equation}
\label{i1}
 \sigma_{\alpha \beta} ( \omega )= \mathrm{Re} \left[ \frac{1}{ i \omega} (K_{\alpha \beta}( \omega+i 0)-K_{\alpha \beta}( 0 ) ) \right],
\end{equation}
where $K_{\alpha \beta}$ represents the thermal Green's function corresponding to the current-current correlation function
\begin{equation}
\label{i2}
 K_{\alpha \beta}( i  \omega_n) = - \frac{e^2}{L^2 \hbar} \int_0^{\hbar/k_\mathrm{B} T} \mathrm{d}\tau e^{i\omega_n \tau} \langle T_\tau j_\alpha (\tau) j_\beta (0) \rangle,
\end{equation}
with $L$ the system size, $T_\tau$ the chronological operator and $\omega_n = 2n\pi k_\mathrm{B} T/\hbar$ for an integer $n$.
The bracket $\langle ... \rangle$ here denotes the ensemble average.
In the calculation of the conductivity tensor~(\ref{i1}), we consider only the loop diagram shown in Fig.\ \ref{Fig-1} and neglect the correction from the current vertex part. The correlation function $K_{\alpha \beta}$ is then written as
\begin{eqnarray}
\label{i3}
\fl K_{\alpha \beta} ( i\omega_n ) =  - \frac{k_\mathrm{B} T e^2}{L^2} \sum_{\omega_m}\sum_{N,k,N',k'} \left<N,k \left| j_\alpha \right|N',k'\right>  \left<N',k' \left| j_\beta \right|N,k\right> \nonumber \\
\times G_{N'} (i\hbar \omega_m + i \hbar \omega_n + \varepsilon_\mathrm{F})  G_{N} ( i \hbar \omega_m + \varepsilon_\mathrm{F}),
\end{eqnarray}
where the electron Green's function $G_N$ is given by Eq.\ (\ref{eq4a}) and the matrix elements of the particle current are 
\begin{eqnarray}
\label{i4}
\fl \left<N,k \left| j_x \right|N',k'\right>  & = &  
 \left( - \frac{\hbar}{m\ell} \sqrt{ \frac{N+1}{2} } \delta_{N',N+1} - 
\frac{\hbar}{m\ell} \sqrt{ \frac{N}{2} } \delta_{N',N-1} \right) \delta_{k,k'},
\nonumber \\
\fl \left<N,k \left| j_y \right|N',k'\right>   & = &  
 \left( - i \frac{\hbar}{m\ell} \sqrt{ \frac{N+1}{2} } \delta_{N',N+1} + i  
\frac{\hbar}{m\ell} \sqrt{ \frac{N}{2} } \delta_{N',N-1} \right) \delta_{k,k'}.
\end{eqnarray}
%
%%%%%%% FIGURE 1 %%%%%%%%%
%%%%%%%%%%%%%%%%%%%%%
%
Performing analytic continuation of $i\omega_n$ to $\omega$ and taking the limit $\omega \rightarrow 0+i\ 0 $, we obtain the dc parts of the diagonal and off-diagonal components in the conductivity tensor in the forms
\begin{eqnarray}
\label{eq4}
\fl  \sigma_{xx} (T,\varepsilon_\mathrm{F})   = \frac{ e^2 }{2 \hbar}(\hbar\omega_c)^2 \int^\infty_{-\infty} \mathrm{d}\varepsilon \left( -\frac{\partial f(\varepsilon)}{\partial\varepsilon}\right) \sum_{N=0}^\infty (N+1) \rho_N(\varepsilon) \rho_{N+1}
(\varepsilon) , 
\\
\label{eq5}
\fl  \sigma_{xy}  (T,\varepsilon_\mathrm{F})  = -\frac{ e^2 }{2\pi\hbar}(\hbar\omega_c)^2 \times \nonumber \\
\fl \hspace{10mm} \sum_{N=0}^\infty \int^\infty_{-\infty} \mathrm{d}\varepsilon 
f(\varepsilon) (N+1) 
   \left( 
 \rho_N(\varepsilon) 
\frac{\partial  G_{N+1}(\varepsilon + i 0) }{\partial \varepsilon} 
 - \rho_{N+1}(\varepsilon) \frac{\partial  G_{N} (\varepsilon + i 0) }{\partial \varepsilon}
  \right) , 
\nonumber \\
 & & 
\end{eqnarray}
where $f(\varepsilon)=1/\{\exp [(\varepsilon-\varepsilon_\mathrm{F})/(k_\mathrm{B}T)] +1 \}$ is the Fermi distribution function. Equations (\ref{eq4}) and (\ref{eq5}) are basically equivalent to Eqs.\ (50) and (51) in Ref.\ \cite{Jonson84} by Jonson and Girvin, except that the self-energy in the Green's function is replaced by a constant value in our case.

We can calculate the components of the conductivity tensor Eqs.\ (\ref{eq4}) and (\ref{eq5}) further using the electron Green's function (\ref{eq4a}). We first examine the diagonal component $\sigma_{xx}$. For $k_\mathrm{B} T \ll \varepsilon_\mathrm{F}$, we can approximate the derivative of the Fermi distribution function by the delta function $-\partial f(\varepsilon)/\partial \varepsilon \simeq \delta(\varepsilon-\varepsilon_\mathrm{F})$. Thus Eq.\ (\ref{eq4}) becomes
\begin{equation}
\label{eq6}
 \sigma_{xx} (\varepsilon_\mathrm{F})  \simeq  \frac{e^2}{2 \hbar}(\hbar\omega_c)^2  \sum_{N=0}^\infty (N+1) \rho_N(\varepsilon_\mathrm{F}) \rho_{N+1} (\varepsilon_\mathrm{F}) .
\end{equation}
Introducing dimensionless parameters
\begin{equation}
X_\mathrm{F}=\frac{\varepsilon_\mathrm{F}}{\hbar \omega_c}-\frac{1}{2},
\end{equation}
and 
\begin{equation}
\gamma=\frac{\Gamma}{\hbar \omega_c},
\end{equation}
we can rewrite Eq.\ (\ref{eq6}) as 
\begin{equation}
\label{eq11}
 \sigma_{xx} (\varepsilon_\mathrm{F})  = \frac{ e^2 }{2\pi^2\hbar} 
 \frac{ {\gamma}^2 }{ 1+4{\gamma}^2 }\sum_{N=0}^\infty 
\frac{ 2X_\mathrm{F} +1 }{ (X_\mathrm{F} - N)^2 + {\gamma}^2}.
\end{equation}
To evaluate the summation over $N$ in Eq.\ (\ref{eq11}), we use the Poisson sum formula
%\begin{equation}
%\label{eq12}
% \sum_{n=-\infty}^{\infty} \alpha \varphi (2\pi n\alpha) = 
%\sum_{\nu=-\infty}^{\infty} \tilde{\varphi} \left( \frac{\nu}{\alpha} %\right)
%\end{equation}
%where $\alpha$ is a real number, $n$ and $\nu$ are integers, and $\tilde{\varphi}$ is the Fourier transform of the function $\varphi$
%\begin{equation}
%\label{eq13}
% \tilde{\varphi}(k)=\frac{1}{2\pi} \int_{-\infty}^\infty \varphi (x) e^{-i k x} \mathrm{d}x.
%\end{equation}
%Applying the Poisson sum formula to the function $\varphi(x)=(x-x_0 \mp i\zeta)^{-1}$, we obtain the relation  
%
\begin{eqnarray}
\sum_{N=-\infty}^\infty \frac{1}{N-X_\mathrm{F} \mp i \gamma} & = & \pm 2\pi i \sum_{\nu=-\infty}^\infty \theta (\mp \nu) e^{-i2\pi\nu X_\mathrm{F} -2\pi | \nu | \gamma}
\nonumber \\ 
 & = & \frac{\pi [-\sin(2\pi X_\mathrm{F}) \pm i \sinh(2\pi\gamma) ]}
{\cosh(2\pi\gamma)-\cos(2\pi X_\mathrm{F})} ,  
\label{eq14}
\end{eqnarray}
%where we used $\tilde{\varphi}(k)$=$\pm i \theta (\mp k) e^{-ik x_0 -|k|\zeta }$ with $\zeta  \rightarrow +0$. 
where $\theta (\xi)$ is the unit step function.
% which equals $0$, $1/2$, and $1$ for $\xi<0$, $\xi=0$ and $\xi>0$, respectively. 
Using Eq.\ (\ref{eq14}), we derive from Eq.\ (\ref{eq11})
\begin{equation}
\label{eq17}
\fl  \sigma_{xx} (\varepsilon_\mathrm{F})  = \frac{ e^2 }{ h } \frac{ 2 {\gamma} }{ 1 + 4 {\gamma}^2}  \frac{( X_\mathrm{F} +1/2 ) \sinh(2\pi {\gamma})}{\cosh(2\pi{\gamma})-\cos(2\pi X_\mathrm{F})} = \frac{ e^2 }{h} \tilde{\sigma}_{xx} (X_\mathrm{F},{\gamma}), 
\end{equation}
where we approximated $\sum\nolimits_{N = 0}^\infty$ by $\sum\nolimits_{N=-\infty}^\infty$, noting that terms with $N<0$ are negligibly small at $\varepsilon_\mathrm{F}$ in the typical situations $\varepsilon_\mathrm{F}$ $\gg$ $\Gamma$. Equation\ (\ref{eq17}) bears the same form as Eq.\ (2.11) in Ref.\ \cite{AndoOC74} by Ando, if we replace our $X_\mathrm{F}$ and $\gamma$ with $X^{\prime}/(\hbar \omega_c)$ and $X^{\prime \prime}/(\hbar \omega_c)$, respectively. In the second equality in Eq.\ (\ref{eq17}), we introduced the notation $\tilde{\sigma}_{\alpha \beta}$ for the conductivity $\sigma_{\alpha \beta}$ normalized by $e^2/h$.

Next we examine the off-diagonal component $\sigma_{xy}$ of the conductivity tensor. Introducing a variable of integration
\begin{equation}
X=\frac{\varepsilon}{\hbar \omega_c}-\frac{1}{2},
\end{equation}
and performing the integration by parts, we rewrite Eq.\ (\ref{eq5}) as
\begin{equation}
\label{eq7}
\fl  \sigma_{xy} (T,\varepsilon_\mathrm{F})  = -\frac{ e^2 }{2\pi\hbar} \sum_{N=0}^\infty (N+1) \int^\infty_{-\infty} \mathrm{d}X \left( -\frac{ \partial f\left(\hbar\omega_c (X+1/2)\right)}{\partial X}  \right) 
L_N ( X ),
\end{equation}   
with 
\begin{eqnarray}
\fl L_N (X) \equiv  (\hbar\omega_c)^2 \int_{-\infty}^X 
  \left[ 
 \rho_N \left(\hbar\omega_c \left( X'+\frac{1}{2} \right) \right) 
\frac{\partial  G_{N+1}\left(\hbar\omega_c (X'+1/2 ) + i 0 \right) }{\partial X'}
 \right. 
\nonumber \\
 \left. -
 \rho_{N+1} \left(\hbar\omega_c \left( X'+\frac{1}{2} \right) \right) 
\frac{\partial  G_{N} \left(\hbar\omega_c (X'+1/2 ) + i 0 \right) }{\partial X'}
\right] \mathrm{d}X'.
\label{eq8}
\end{eqnarray}
For $k_\mathrm{B}T \ll \varepsilon_\mathrm{F}$, we obtain
\begin{equation}
\label{eq9}
 \sigma_{xy} (\varepsilon_\mathrm{F})  = -\frac{ e^2 }{h} \sum_{N=0}^\infty (N+1) L_N ( X_\mathrm{F} ).
\end{equation}
Using Eq.\ (\ref{eq4a}) and performing the integration $L (X)$ in Eq.\ (\ref{eq8}) up to $X_\mathrm{F}$, we obtain
\begin{eqnarray}
\label{eq10}
\fl  \sigma_{xy} (\varepsilon_\mathrm{F})  = -\frac{ e^2 }{2\pi^2\hbar}\left\{ 
 \frac{ 2{\gamma}^3 }{ 1+4{\gamma}^2 }\sum_{N=0}^\infty 
\left[
\frac{1}{2 {\gamma}^2 }\frac{ (X_\mathrm{F}-N) }{ (X_\mathrm{F} - N)^2 + {\gamma}^2 } - 
\frac{2N+1}{(X_\mathrm{F} - N)^2 + {\gamma}^2 }
\right] \right. 
\nonumber 
\\
\left. +\sum_{N=0}^\infty \left[ \arctan\left( \frac{X_\mathrm{F}-N}{{\gamma}} \right) + \frac{\pi}{2}  \right] \right\} .
\end{eqnarray}   
We can evaluate the summation over $N$ in Eq.\ (\ref{eq10}), following a similar procedure as in the calculation from Eqs.\ (\ref{eq11}) to (\ref{eq17}). The first line in the right hand side (r.h.s.) of Eq.\ (\ref{eq10}) becomes
\begin{equation}
\fl \frac{e^2}{h}
\frac{1}{ \cosh(2\pi\gamma)-\cos(2\pi X_\mathrm{F}) } 
\left[
 -{\gamma} \sin(2\pi X_\mathrm{F}) + 
\frac{ 4 {\gamma}^2 ( X_\mathrm{F} +1/2 ) }{1+4{\gamma}^2}  \sinh(2\pi{\gamma})
\right],
\label{eq15-1}
\end{equation}
where we used Eq.\ (\ref{eq14}), employing the approximation  $\sum\nolimits_{N = 0}^\infty$ $\rightarrow$  $\sum\nolimits_{N=-\infty}^\infty$ as before. Along the same line, we can accurately approximate the last term in the r.h.s.\ of Eq.\ (\ref{eq10}) by
\begin{equation}
  - \frac{ e^2 }{h} \frac{1}{\pi}
  \left[ 
\sum_{N=-\infty}^\infty 
\arctan\left( \frac{X_\mathrm{F} - N}{{\gamma}}  
\right) +\frac{\pi}{2} 
\right] ,
\label{eq15-2}
\end{equation}
noting that $\arctan ((X_\mathrm{F}-N)/{\gamma}) \simeq \pi/2$ for $N<0$ since $(X_\mathrm{F}-N)/{\gamma}$ = $[\varepsilon_\mathrm{F}-(N+1/2)\hbar \omega_c]/\Gamma \gg 0$ for $\varepsilon_\mathrm{F} \gg \Gamma$. Using the relation $ \arctan ( (X-N) / {\gamma} )  = \int {\gamma}/[ (X-N)^2 + {\gamma}^2 ] \mathrm{d} X + \mbox{const.}$ 
 and Eq.\ (\ref{eq14}), we can rewrite Eq.\ (\ref{eq15-2}) further as
\begin{eqnarray}
- \frac{ e^2 }{h} 
  \left[ 
\sum_{N=-\infty}^{\infty} 
     \int_{0}^{X_\mathrm{F}} \frac{1}{\pi}\frac{ {\gamma}}{ (X-N)^2 + {\gamma}^2 }\mathrm{d}X 
 + \frac{1}{2}
\right] 
 \nonumber \\
= - \frac{ e^2 }{h} 
  \left[ 
\frac{1}{\pi} \arctan \left( \coth(\pi{\gamma})\tan(\pi X_\mathrm{F} ) \right)+ \mbox{\rm Int}\left(  X_\mathrm{F}+\frac{1}{2} \right)  + \frac{1}{2}
\right] ,
\label{eq15-3}
\end{eqnarray}
with $\mathrm{Int}(\xi)$ representing the integer part of $\xi$. We finally arrive at
\begin{eqnarray}
\label{eq16}
\fl \sigma_{xy} (\varepsilon_\mathrm{F})  = \frac{ e^2 }{h}
 \left\{
\frac{1}{ \cosh(2\pi\gamma)-\cos(2\pi X_\mathrm{F}) } 
\left[
\frac{ 4 {\gamma}^2 ( X_\mathrm{F} +1/2 )}{1+4{\gamma}^2}  \sinh(2\pi{\gamma})
 -{\gamma} \sin(2\pi X_\mathrm{F}) \right] \right.
\nonumber 
\\
\left. - 
\frac{1}{\pi} \arctan \left( \coth(\pi{\gamma})\tan(\pi X_\mathrm{F} ) \right) - \mbox{\rm Int}\left(  X_\mathrm{F}+\frac{1}{2} \right)  - \frac{1}{2}
\right\}  
\nonumber \\
= \frac{ e^2 }{h} \tilde{\sigma}_{xy} (X_\mathrm{F},{\gamma}) .
\end{eqnarray}
%
%Equation (\ref{eq16}) is one of our main results. 
As far as we know, an explicit analytic formula for $\sigma_{xy}$ has never been reported thus far.

%%%%%%% FIGURE 2 %%%%%%%%%
\begin{figure}[tb]
\begin{center}
\includegraphics[width=8.5cm]{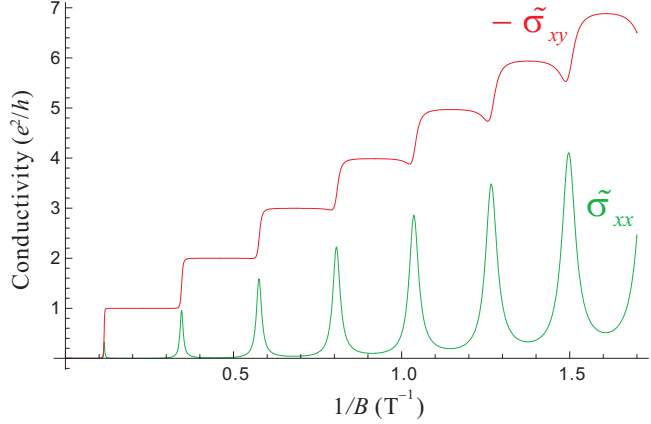}
\end{center}
\caption{\label{Fig-2}The diagonal [Eq.\ (\ref{eq20a})] and the off-diagonal [Eq.\ (\ref{eq20b})] components of the conductivity tensor. The horizontal axis is the inverse magnetic field. }
\end{figure}
%%%%%%%%%%%%%%%%%%%%%

In Fig.\ \ref{Fig-2}, we show the diagonal $\tilde\sigma_{xx}$ and off-diagonal $\tilde\sigma_{xy}$ components of the normalized conductivity tensor calculated by Eqs.\ (\ref{eq17}) and (\ref{eq16}) [or equivalently, by Eqs.\ (\ref{eq20a}) and (\ref{eq20b}) below], respectively.  The parameters are selected to be typical values in a GaAs/AlGaAs 2DES: $m^*=0.067 m_0$ with $m_0$ the bare electron mass, $\varepsilon_\mathrm{F}=7.5$ meV, and $\Gamma=\hbar/(2\tau)=0.12$ meV. The traces basically reproduce well-known behavior of a 2DES in the magnetic field: the staircase with plateaus at integer multiples of $e^2/h$ for $\sigma_{xy}$ and peaks at inter-plateau transition for $\sigma_{xx}$. The non-monotonic $1/B$ dependence observed in $\sigma_{xy}$ for $B \leq $ 1 T (the depression in $-\sigma_{xy}$ that occurs in step with the peak in $\sigma_{xx}$) is usually not seen in the experimental traces for a high-mobility GaAs/AlGaAs 2DES, but can be seen in early experiments on Si-MOSFET \cite{Wakabayashi78} and is likely to be related to the short-range nature of the impurity potential. (See Fig.\ \ref{sgmxxsgmxytr} below for comparison with the result in the long-range potential.)

For brevity and for the convenience in later use, we rewrite $\tilde\sigma_{xx}$ and $\tilde\sigma_{xy}$ in concise formulas,
\begin{eqnarray}
 \label{eq20a}
\fl \tilde{\sigma}_{xx} (X_\mathrm{F},{\gamma})  = \frac{2 \gamma}{1+4\gamma^2} \left( X_\mathrm{F}+\frac{1}{2} \right) {\rm Fsinh} (X_\mathrm{F},\gamma)
\\
 \label{eq20b}
\fl \tilde{\sigma}_{xy} (X_\mathrm{F},{\gamma})  =  - {\rm IFsinh}(X_\mathrm{F},\gamma)-{\gamma} {\rm Fsin} (X_\mathrm{F},\gamma) + \frac{4\gamma^2}{1+4\gamma^2} \left( X_\mathrm{F}+\frac{1}{2} \right) {\rm Fsinh} (X_\mathrm{F},\gamma), \nonumber \\
\end{eqnarray}  
where we introduced the notations ${\rm Fsin}(X_\mathrm{F},{\gamma})$, ${\rm Fsinh}(X_\mathrm{F},{\gamma})$, and ${\rm IFsinh} (X_\mathrm{F},{\gamma})$ defined as
\begin{eqnarray}
\label{eq20}
\fl {\rm Fsin}(X_\mathrm{F},{\gamma})  & =  \frac{\sin (2\pi X_\mathrm{F})}{\cosh(2\pi{\gamma})-\cos (2\pi X_\mathrm{F})},
\nonumber \\
\fl {\rm Fsinh}(X_\mathrm{F},{\gamma})  & =  \frac{\sinh (2\pi \gamma)}{\cosh(2\pi{\gamma})-\cos (2\pi X_\mathrm{F})},
\nonumber \\
\fl {\rm IFsinh} (X_\mathrm{F},{\gamma}) &  =  \int_{-\frac{1}{2}}^{X_\mathrm{F}} \mathrm{d}X {\rm Fsinh}(X,{\gamma}) \nonumber \\
& = \frac{1}{\pi}\arctan \left(  \coth(\pi{\gamma}) \tan(\pi X_\mathrm{F} ) \right) + \mbox{\rm Int}\left(  X_\mathrm{F}+\frac{1}{2} \right)  + \frac{1}{2} . 
\end{eqnarray}
Although it appears, at first glance, that the stepwise behavior of $\sigma_{xy}$ is reflecting only the first term in Eq.\ (\ref{eq20b}), the second term is also playing its own share of roles by extending the width of the plateau and thus making the slope of the inter-plateau region much steeper than it would be were it not for the term. The steepness of the slope is of paramount importance in our theory that attempts to explain the behavior of $\mathrm{d} \sigma_{xy}/\mathrm{d}B$.

We note in passing that the DOS given by Eq.\ (\ref{eq2a}) can also be rewritten, following the same procedure as in the derivation of Eq.\ (\ref{eq20a}), as
\begin{equation}
D(\varepsilon) = D_0 \mathrm{Fsinh}(X,\gamma), \label{DOSF}
\end{equation}
where $D_0$ = $m^* / (2 \pi \hbar^2)$ represents the DOS of a 2DES in the absence of the magnetic field, and $X=\varepsilon/(\hbar \omega_c)-1/2$ as defined earlier. Accordingly, the cumulative number of states $N(\varepsilon)$ below $\varepsilon$ reads
\begin{equation}
N(\varepsilon) = \int _0^\varepsilon {D(\varepsilon^\prime) \mathrm{d}\varepsilon^\prime} = \frac{1}{2 \pi \ell^2} \mathrm{IFsinh}(X,\gamma). \label{NEF}
\end{equation}

We will show in \ref{WeakMag} that Eqs.\ (\ref{eq20a}) and (\ref{eq20b}) tends to the well-known semiclassical formulas for $B \rightarrow 0$.

\section{The relation between diagonal and off-diagonal conductivities at high magnetic fields\label{Relation}} 
We now move on to the main topic of the present paper, the relation between $\tilde{\sigma}_{xx}$ and $\tilde{\sigma}_{xy}$ at high magnetic fields.  Since both of $X_\mathrm{F}$ and $\gamma$ are functions of $B$, the derivative of the off-diagonal component $\tilde{\sigma}_{xy}$ with respect to $B$ is written as
\begin{eqnarray}
\label{eq18}
\fl \frac{\mathrm{d} \tilde{\sigma}_{xy}(X_\mathrm{F},{\gamma})}{\mathrm{d} B} &= \frac{\partial \tilde{\sigma}_{xy}(X_\mathrm{F},{\gamma})}{\partial X_\mathrm{F}} \frac{ \mathrm{d} X_\mathrm{F}}{\mathrm{d} B} + \frac{\partial \tilde{\sigma}_{xy}(X_\mathrm{F},{\gamma})}{\partial {\gamma}} \frac{\mathrm{d} {\gamma} }{\mathrm{d} B} \nonumber \\
 &= -\frac{1}{B}\left[ \left( X_\mathrm{F}+\frac{1}{2} \right) \frac{\partial \tilde{\sigma}_{xy}(X_\mathrm{F},{\gamma}) }{\partial X_\mathrm{F}} + {\gamma}\frac{\partial \tilde{\sigma}_{xy} (X_\mathrm{F},{\gamma})}{\partial {\gamma}}  \right].
\end{eqnarray} 
Differentiation by $X_\mathrm{F}$ and by $\gamma$ can be analytically done on Eq.\ (\ref{eq20b}) and we obtain
\begin{eqnarray}
\label{eq19}
\fl B \frac{\mathrm{d} \tilde{\sigma}_{xy}(X_\mathrm{F},{\gamma})}{\mathrm{d} B} & = & \left( X_\mathrm{F} + \frac{1}{2} \right) 
\left[ \frac{1-4{\gamma}^2}{(1+4{\gamma}^2)^2} -\frac{1+8{\gamma}^2}{1+4{\gamma}^2}2\pi {\gamma} \coth(2\pi {\gamma}) \right]
{\rm Fsinh}(X_\mathrm{F}, {\gamma})
\nonumber \\
 & & + \left( X_\mathrm{F} +\frac{1}{2} \right) \frac{1+8{\gamma}^2}{1+4{\gamma}^2} 2 \pi {\gamma} \mathrm{Fsinh}^2(X_\mathrm{F},{\gamma})
\nonumber \\
 & & - \left[ 1-\frac{4}{1+ 4 {\gamma}^2 } \left( X_\mathrm{F}+\frac{1}{2} \right)^2 \right] 2 \pi {\gamma}^2 {\rm Fsin(X_\mathrm{F},{\gamma})} {\rm Fsinh} (X_\mathrm{F},{\gamma}),
\end{eqnarray}
or, with the aid of Eq.\ (\ref{eq20a}),
\begin{eqnarray}
\label{eq21}
\fl B \frac{\mathrm{d} \tilde{\sigma}_{xy}(X_\mathrm{F},{\gamma})}{\mathrm{d} B} & = &  
\frac{1}{2\gamma}\left[ \frac{1-4{\gamma}^2}{1+4{\gamma}^2} - 2\pi \gamma (1+8{\gamma}^2) {\rm coth} (2\pi {\gamma}) \right]\tilde{\sigma}_{xx}(X_\mathrm{F},{\gamma})
\nonumber \\
 & &  + \frac{ \pi }{ 2\gamma } \frac{(1+8{\gamma}^2)(1+4{\gamma}^2)}{X_\mathrm{F} + 1/2}  \tilde{\sigma}_{xx}^2 (X_\mathrm{F},{\gamma})
\nonumber \\
 & & - \frac{ \pi ( 1 + 4 {\gamma}^2 ) }{ 2\sinh(2\pi\gamma) } \left[ \frac{ 1+ 4 {\gamma}^2}{ \left( X_\mathrm{F}+1/2 \right)^2 } -4 \right] \sin(2\pi X_\mathrm{F})\tilde{\sigma}_{xx}^2 (X_\mathrm{F},{\gamma}). 
\end{eqnarray}

\begin{figure}[tb]
\begin{center}
\includegraphics[width=8.5cm]{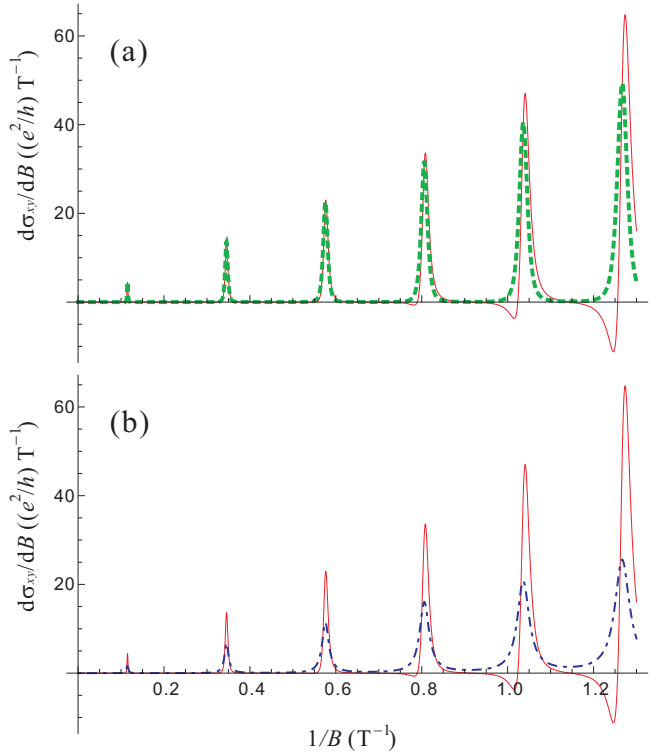}
\end{center}
\caption{\label{Fig-3}The plots of $\mathrm{d}\tilde{\sigma}_{xy}/\mathrm{d}B$ calculated by an unabridged equation, Eq.\ (\ref{eq19}) [thin solid red line, plotted in both (a) and (b)], and by an approximated equation, Eq.\ (\ref{eq22}), with $\tilde{\sigma}_{xx}$ calculated by Eq.\ (\ref{eq20a}) [thick dashed green line in (a)]. We also plot $\mathrm{d}\tilde{\sigma}_{xy}/\mathrm{d}B^{(1)}$ in Eq.\ (\ref{wrongapprox}), obtained by keeping only the first term in Eq.\ (\ref{eq20b}), for comparison [dot-dashed blue line in (b)]. The traces are separately plotted in (a) and/or (b) for clarity.}
\end{figure}

We will pick out the dominant term at high magnetic fields from the r.h.s.\ of Eq.\ (\ref{eq21}). Since ${\gamma}=\Gamma/(\hbar \omega_c)$ tends to zero with the increase of the magnetic field, we expand the coefficients in terms of ${\gamma}$ for this purpose as
\begin{eqnarray}
\label{eq21a}
\fl B \frac{\mathrm{d} \tilde{\sigma}_{xy}(X_\mathrm{F},{\gamma})}{\mathrm{d} B} & = &  
\left[ -\left( 8+\frac{2 \pi^2}{3} \right) \gamma +O(\gamma^2) \right] \tilde{\sigma}_{xx}(X_\mathrm{F},{\gamma})
\nonumber \\
 & &  + \left[ \frac{\pi}{2 \gamma \left( X_\mathrm{F} + \frac{1}{2} \right)} + O(\gamma) \right]  \tilde{\sigma}_{xx}^2 (X_\mathrm{F},{\gamma})
\nonumber \\
 & & - \left\{ \frac{1}{\gamma} \left[ \frac{ 1}{ 4 \left( X_\mathrm{F}+\frac{1}{2} \right)^2 } -1 \right] +O(\gamma) \right\} \sin(2\pi X_\mathrm{F})\tilde{\sigma}_{xx}^2 (X_\mathrm{F},{\gamma}).
\end{eqnarray}
The diagonal component $\tilde{\sigma}_{xx}$ can be readily seen from Eq.\ (\ref{eq20a}) to take peaks at $X_\mathrm{F} = N$ (integer), namely when the Fermi energy lies at the center of $N$th Landau level, with the peak height given by
\begin{eqnarray}
\fl \tilde{\sigma}_{xx}(N,{\gamma}) & = & \frac{2 \gamma}{1+4\gamma^2} \left( X_\mathrm{F}+\frac{1}{2} \right) \frac{\sinh(2\pi \gamma)}{\cosh(2\pi \gamma)-1} \nonumber \\
\fl & = & \left( X_\mathrm{F}+\frac{1}{2} \right) \left[ \frac{2}{\pi}+\left(-\frac{8}{\pi}+\frac{2 \pi}{3} \right) \gamma^2 + O(\gamma^4) \right],
\label{sgmxxexp}
\end{eqnarray}
and $\tilde{\sigma}_{xx} \sim 0$ away from the sharp peaks (see also Fig.\ \ref{Fig-2}). From Eqs.\ (\ref{eq21a}) and (\ref{sgmxxexp}), and noting that $\sin(2 \pi X_\mathrm{F}) \sim 0$ at $X_\mathrm{F} \sim N$, we find that the second term in Eq.\ (\ref{eq21a}) makes the dominant contribution, leading to our final result,
\begin{equation}
\label{eq22}
\frac{\mathrm{d} \tilde{\sigma}_{xy}(X_\mathrm{F},{\gamma})}{\mathrm{d} B} \simeq  \pi \mu \frac{\hbar \omega_c}{\varepsilon_\mathrm{F}} \tilde{\sigma}_{xx}^2 (X_\mathrm{F},{\gamma}),
\end{equation}
or $\lambda = (h/e^2) \pi \hbar e \mu (m^* \varepsilon_\mathrm{F})^{-1}$ in Eq.\ (\ref{FR1}).
Here we made use of the mobility $\mu = e\tau/m^* = e\hbar / (2 m^* \Gamma)$. Plots of $\mathrm{d}\tilde{\sigma}_{xy}/\mathrm{d}B$ calculated using Eq.\ (\ref{eq19}) (solid red line) and $\pi \mu (\hbar \omega_c/\varepsilon_\mathrm{F}) \tilde{\sigma}_{xx}^2$ with $\tilde{\sigma}_{xx}$ computed by Eq.\ (\ref{eq20a}) (dashed green line) shown in Fig.\ \ref{Fig-3} attest to the validity of Eq.\ (\ref{eq22}) for $B \geq 1$ T\@. The deviation seen at lower magnetic fields is attributable to higher order terms in $\gamma$ neglected in Eq.\ (\ref{eq22}). In Fig.\ \ref{Fig-3}, we used the same parameter values as in Fig.\ \ref{Fig-2}.

%%%%%%% FIGURE 3 %%%%%%%%%
%%%%%%%%%%%%%%%%%%%%% 

It is interesting to point out that we obtain the relation $\mathrm{d} \sigma_{xy}/\mathrm{d}B \propto \sigma_{xx}$ instead of Eq.\ (\ref{eq22}) if we keep only the first term in Eq.\ (\ref{eq20b}),
\begin{equation}
\label{wrongapprox}
\fl \frac{\mathrm{d} \tilde{\sigma}_{xy}(X_\mathrm{F},{\gamma})}{\mathrm{d} B}^{(1)} = \frac{1}{B} \left[\left(X_\mathrm{F}+\frac{1}{2}\right) \mathrm{Fsinh}(X_\mathrm{F},\gamma)-\gamma  \mathrm{Fsin}(X_\mathrm{F},\gamma )\right] \simeq \mu \tilde{\sigma}_{xx}, 
\end{equation}
which is not legitimate as discussed below Eq.\ (\ref{eq20}) in Sec.\ \ref{CondTen}. In fact, the peaks calculated by Eq.\ (\ref{wrongapprox}) exhibit much larger width and smaller (roughly half) height compared with those calculated by Eq.\ (\ref{eq19}), as displayed in Fig.\ \ref{Fig-3}.

\section{Comparison with experimental results\label{CompExp}}
\subsection{Modification for long-range potential}
In this section, we make an attempt to compare the relation between $\tilde{\sigma}_{xx}$ and $\tilde{\sigma}_{xy}$ deduced in Sec.\ \ref{Relation} to the experimental results obtained in a GaAs/AlGaAs 2DES\@. It is well known that the main source of scattering in a GaAs/AlGaAs 2DES is the ionized donors. The donors are set back from the 2DES plane by a spacer layer with the thickness typically a few tens of nanometers. Therefore, the scattering in a GaAs/AlGaAs 2DES should be described by a long-range impurity potential. Since a short-range potential is assumed in our theory, slight modification is necessary to implement the comparison. This is done by following the prescription given in Ref.\ \cite{Coleridge89}.

First we observe that the off-diagonal component Eq.\ (\ref{eq20b}) can be rewritten in the form presented in Ref.\ \cite{Ishihara86},
$\sigma_{xy} = -e [\partial N(\varepsilon_\mathrm{F}) / \partial B] - \omega_c \tau \sigma_{xx}$,
which reads in the normalized form,
\begin{equation}
\tilde{\sigma}_{xy} = -\frac{h}{e} \frac{\partial N(\varepsilon_\mathrm{F})}{\partial B}-\frac{1}{2 \gamma} \tilde{\sigma}_{xx}. \label{IandS}
\end{equation}
The equivalence of Eq.\ (\ref{IandS}) to Eq.\ (\ref{eq20b}) can readily be verified by performing the differentiation by $B$ on $N(\varepsilon_\mathrm{F})$ given by Eq.\ (\ref{NEF}):
\begin{equation}
\fl \frac{h}{e} \frac{\partial N(\varepsilon_\mathrm{F})}{\partial B} = \mathrm{IFsinh}(X_\mathrm{F},\gamma) + \gamma \mathrm{Fsin}(X_\mathrm{F},\gamma ) - \left( X_\mathrm{F} + \frac{1}{2} \right) \mathrm{Fsinh}(X_\mathrm{F},\gamma ).  
\end{equation}

In a long-range potential, it is important to recall that the scattering is characterized by two distinct scattering times, namely, the quantum scattering time $\tau_q$ =$\hbar/(2\Gamma)$ that describes the impurity broadening of the Landau levels and the momentum relaxation time $\tau_m=\sigma_0 m^*/(n_e e^2)$ related to the conductivity at $B=0$. The latter time is typically 10 times larger than the former in a GaAs/AlGaAs 2DES, while the relaxation times are simply $\tau_q=\tau_m=\tau$ for short-range scatterers. Coleridge \textit{et al.} \cite{Coleridge89} suggested an appropriate way of replacing $\tau$ by either of $\tau_q$ or $\tau_m$, with which the resultant $\sigma_{xx}$ and $\sigma_{xy}$ describe the conductivities under the long-range potential quite well. The method, in our notation, is to replace $\gamma$ only in the prefactor of Eq.\ (\ref{eq20a}) by $\gamma_m=1/(2 \omega_c \tau_m)$, leaving $\gamma=\gamma_q=\Gamma/(\hbar \omega_c)=1/(2 \omega_c \tau_q)$ in $\mathrm{Fsinh}(X,\gamma)$ intact:
\begin{equation}
\tilde{\sigma}^\mathrm{LR}_{xx} (X_\mathrm{F},\gamma_q,\gamma_m) = \frac{2 \gamma_m}{1+4\gamma_m^2} \left( X_\mathrm{F}+\frac{1}{2} \right) {\rm Fsinh} (X_\mathrm{F},\gamma_q). \label{sgmxxLR}
\end{equation}
The Hall conductivity is obtained by substituting $\gamma_m$ and  $\tilde{\sigma}^\mathrm{LR}_{xx}$ into the second term of Eq.\ (\ref{IandS}) as
\begin{eqnarray}
\fl \tilde{\sigma}^\mathrm{LR}_{xy} (X_\mathrm{F},\gamma_q,\gamma_m) & = -\left. \frac{h}{e} \frac{\partial N(\varepsilon_\mathrm{F})}{\partial B} \right|_{\gamma = \gamma_q} -\frac{1}{2 \gamma_m} \tilde{\sigma}^\mathrm{LR}_{xx} \nonumber\\
 & = - {\rm IFsinh}(X_\mathrm{F},\gamma_q)-{\gamma_q} {\rm Fsin} (X_\mathrm{F},\gamma_q) + 2 \gamma_m \tilde{\sigma}^\mathrm{LR}_{xx}. \label{sgmxyLR}
\end{eqnarray}
With these substitutions, the derivative of $\tilde{\sigma}^\mathrm{LR}_{xy}$ by $B$ reads
\begin{equation}
\fl \frac{\mathrm{d} \tilde{\sigma}^\mathrm{LR}_{xy}(X_\mathrm{F},{\gamma_q},\gamma_m)}{\mathrm{d} B} = \eta_1 (X_\mathrm{F},\gamma_q) + \eta_2 (X_\mathrm{F},\gamma_q) + \eta_3 (X_\mathrm{F},\gamma_q,\gamma_m)
\label{fundrelwoapp} 
\end{equation}
with
\begin{displaymath}
\fl \eta_1 (X_\mathrm{F},\gamma_q) = \frac{1}{B} \left[\left(X_\mathrm{F}+\frac{1}{2}\right) \mathrm{Fsinh}(X_\mathrm{F},\gamma_q )-\gamma_q  \mathrm{Fsin}(X_\mathrm{F},\gamma_q )\right], 
\end{displaymath}
\begin{eqnarray}
\fl \eta_2 (X_\mathrm{F},\gamma_q) = \frac{1}{B} \left\{\left(X_\mathrm{F}+\frac{1}{2}\right) 2 \pi  \gamma_q \mathrm{Fsinh}(X_\mathrm{F},\gamma_q ) \left[   \mathrm{Fsinh}(X_\mathrm{F},\gamma_q )- \mathrm{coth}(2 \pi  \gamma_q ) \right] \right. \nonumber \\
\biggl. + \gamma_q  \mathrm{Fsin}(X_\mathrm{F},\gamma_q )-2 \pi  \gamma_q ^2 \mathrm{Fsin}(X_\mathrm{F},\gamma_q ) \mathrm{Finh}(X_\mathrm{F},\gamma_q )\biggr\}, \nonumber
\end{eqnarray}
and
\begin{eqnarray}
\fl \eta_3 (X_\mathrm{F},\gamma_q,\gamma_m) = -\frac{2 \gamma_m}{B} \tilde{\sigma}_{xx}^\mathrm{LR}(X_\mathrm{F},\gamma_q ,\gamma_m ) \nonumber \\
\fl \times \left\{ 2 \pi \gamma_q  \mathrm{coth}(2 \pi  \gamma_q )+1+\frac{2}{1+4 \mathrm{\gamma_m }^2} -2 \pi  \left[ \left(X_\mathrm{F}+\frac{1}{2}\right) \mathrm{Fsin}(X_\mathrm{F},\gamma )+\gamma_q  \mathrm{Fsinh}(X_\mathrm{F},\gamma_q )\right] \right\}, \nonumber
\end{eqnarray}
where the terms $\eta_1$, $\eta_2$ and $\eta_3$ are derived from the first, the second and the third term in Eq.\ (\ref{sgmxyLR}), respectively.
Accordingly, the dominant term at high magnetic field changes from Eq.\ (\ref{eq22}) to
\begin{equation}
\frac{\mathrm{d} \tilde{\sigma}^\mathrm{LR}_{xy}(X_\mathrm{F},{\gamma_q},\gamma_m)}{\mathrm{d} B} \simeq  \pi \frac{\mu_m^2}{\mu_q} \frac{\hbar \omega_c}{\varepsilon_\mathrm{F}} \left[\tilde{\sigma}^\mathrm{LR}_{xx} (X_\mathrm{F},{\gamma_q},\gamma_m) \right]^2, \label{fundrelLR}
\end{equation}
[$\lambda = (h/e^2) \pi \hbar e (\mu_m^2/\mu_q) (m^* \varepsilon_\mathrm{F})^{-1}$ in Eq.\ (\ref{FR1})], where $\mu_q$ = $e \tau_q/m^*$ and $\mu_m$ = $e \tau_m/m^*$ are mobilities corresponding to $\tau_q$ and $\tau_m$, respectively.

%%%%%%% FIGURE 4 %%%%%%%%%
\begin{figure}[tb]
\begin{center}
\includegraphics[width=8.5cm]{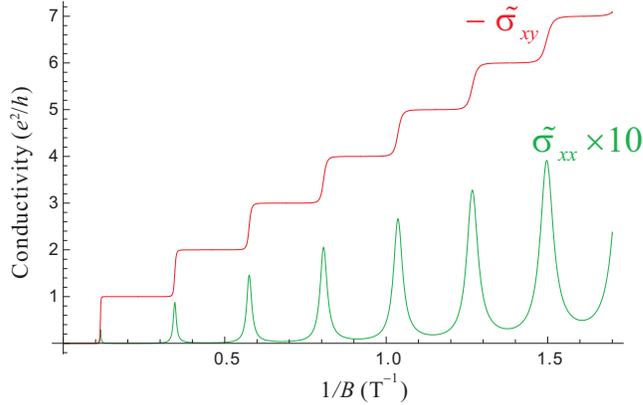}
\end{center}
\caption{\label{sgmxxsgmxytr} The diagonal [Eq.\ (\ref{sgmxxLR})] and the off-diagonal [Eq.\ (\ref{sgmxyLR})] components of the conductivity tensor modified to account for the long-range potential. The horizontal axis is the inverse magnetic field. }
\end{figure}
%%%%%%%%%%%%%%%%%%%%%

%%%%%%% FIGURE 5 %%%%%%%%%
\begin{figure}[tb]
\begin{center}
\includegraphics[width=8.5cm]{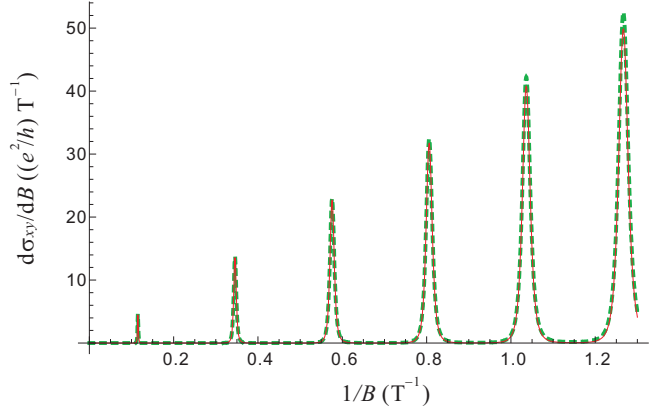}
\end{center}
\caption{\label{fundreltr} The plots of $\mathrm{d}\tilde{\sigma}^\mathrm{LR}_{xy}/\mathrm{d}B$ without approximation, Eq.\ (\ref{fundrelwoapp}) (thin solid red line), and $\mathrm{d}\tilde{\sigma}^\mathrm{LR}_{xy}/\mathrm{d}B$ approximated by Eq.\ (\ref{fundrelLR}) with $\tilde{\sigma}^\mathrm{LR}_{xx}$ calculated by Eq.\ (\ref{sgmxxLR}) (thick dashed green line). }
\end{figure}
%%%%%%%%%%%%%%%%%%%%%

In Fig.\ \ref{sgmxxsgmxytr}, we show the longitudinal and the Hall conductivities calculated by Eqs.\ (\ref{sgmxxLR}) and (\ref{sgmxyLR}) with parameters $\varepsilon_\mathrm{F}$ = 7.5 meV, $\mu_q$ = 7.1 m$^2$/(Vs) (corresponding to $\Gamma$ = 0.12 meV), and $\mu_m$ = 78 m$^2$/(Vs). The parameters are taken from our experiment to be presented below. The diagonal component $\sigma_{xx}$ has become much smaller than in Fig.\ \ref{Fig-2} (note the 10 times magnification in Fig.\ \ref{sgmxxsgmxytr}), in accordance with experiments in a GaAs/AlGaAs 2DES\@. Note that the non-monotonic behavior of $\sigma_{xy}$ observed in Fig.\ \ref{Fig-2} has vanished in Fig.\ \ref{sgmxxsgmxytr}. The high accuracy of the approximation given by Eq.\ (\ref{fundrelLR}) at high enough magnetic fields ($B \geq $ 1 T) is demonstrated in Fig.\ \ref{fundreltr}. 

%%%%%%% FIGURE 6 %%%%%%%%%
\begin{figure}[tb]
\begin{center}
\includegraphics[width=8.5cm]{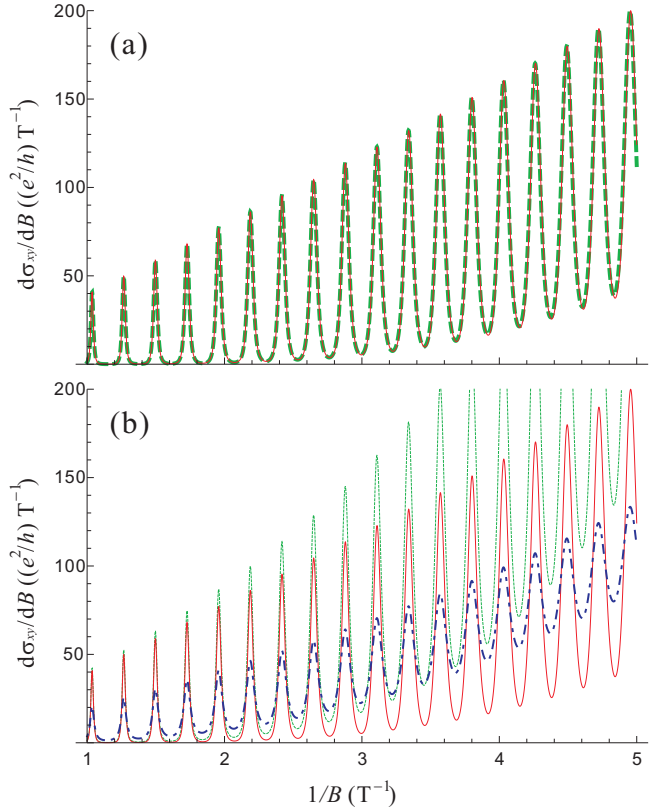}
\end{center}
\caption{\label{fundrelLF} The plots of $\mathrm{d}\tilde{\sigma}^\mathrm{LR}_{xy}/\mathrm{d}B$ without approximation, Eq.\ (\ref{fundrelwoapp}) [thin solid red line, plotted in both (a) and (b)], and $\mathrm{d}\tilde{\sigma}^\mathrm{LR}_{xy}/\mathrm{d}B$ approximated by Eq.\ (\ref{fundrelLRn}) [thick dashed green line in (a)] or by Eq.\ (\ref{fundrelLR}) [thin dotted green line in (b)] for a lower magnetic field range than in Fig.\ \ref{fundreltr}. We also plot $\mathrm{d}\tilde{\sigma}_{xy}/\mathrm{d}B^{(1)}$ in Eq.\ (\ref{wrongapprox}) for comparison [dot-dashed blue line in (b)]. The traces are separately plotted in (a) and/or (b) for clarity.}
\end{figure}
%%%%%%%%%%%%%%%%%%%%%

Although Eq.\ (\ref{fundrelLR}) as well as Eq.\ (\ref{eq22}) is intended for the use in high magnetic fields, stringent comparison with experimental results is possible only in a rather low magnetic-field range ($B \leq  $0.5 T) for a couple of reasons to be discussed in Sec.\ \ref{Disc}. Above all, we neglected the spin of the electrons altogether in the theory. In principle, the spin can be included in the theory by adding $\sigma g \mu_\mathrm{B} B$ to $E_N$ with $\sigma = \pm1/2$ representing the spin and $\mu_\mathrm{B}$ the Bohr magneton. Difficulty arises, however, because of the dependence of the $g$-factor on the magnetic field owing to the exchange interaction \cite{AndoOG74}: the $g$-factor experiences strong enhancement at the magnetic field where the Fermi energy lies between the Zeeman gap (exchange enhancement), which defies simple analytical treatment. If we limit ourselves to $B \leq $0.5 T, spin splitting can be completely neglected because of the small (bare) $g$-factor $g = -0.44$ in GaAs. In this low magnetic field range, the approximation in Eq.\ (\ref{fundrelLR}) that retains only the leading term in $\gamma$ turns out to be insufficient, as demonstrated in Fig.\ \ref{fundrelLF}. The approximation is improved by keeping the terms deriving from the first two terms in the r.h.s.\ of Eq.\ (\ref{sgmxyLR}), $\eta_1$ and $\eta_2$, except for the term including $\mathrm{Fsin}(X_\mathrm{F},\gamma)$ (the third term $\eta_3$ can safely be neglected since $\gamma_m \ll \gamma_q$):
\begin{eqnarray}
\fl \frac{\mathrm{d} \tilde{\sigma}^\mathrm{LR}_{xy}(X_\mathrm{F},{\gamma_q},\gamma_m)}{\mathrm{d} B} \simeq & \pi \frac{\mu_m^2}{\mu_q} \frac{\hbar \omega_c}{\varepsilon_\mathrm{F}} \left[\tilde{\sigma}^\mathrm{LR}_{xx} (X_\mathrm{F},{\gamma_q},\gamma_m) \right]^2 \nonumber \\ & + \mu_m \left[ 1-\frac{\pi}{\mu_q B} \coth \left( \frac{\pi}{\mu_q B} \right) \right] \tilde{\sigma}^\mathrm{LR}_{xx} (X_\mathrm{F},{\gamma_q},\gamma_m). \label{fundrelLRn}
\end{eqnarray}
In Fig.\ \ref{fundrelLF}, we plot $\mathrm{d} \tilde{\sigma}^\mathrm{LR}_{xy}(X_\mathrm{F},{\gamma})/\mathrm{d} B$ without approximation, Eq.\ (\ref{fundrelwoapp}), along with approximated traces, Eqs.\ (\ref{fundrelLR}) and (\ref{fundrelLRn}), calculated using $\tilde{\sigma}^\mathrm{LR}_{xx}(X_\mathrm{F},{\gamma})$ in Eq.\ (\ref{sgmxxLR}). Deviation of Eq.\ (\ref{fundrelLR}) from the exact result becomes evident below $\sim$0.5 T, while Eq.\ (\ref{fundrelLRn}) reproduces the trace almost indistinguishable from that of the exact calculation in the magnetic field range shown in Fig.\ \ref{fundrelLF}.

\subsection{Relation between experimentally observed longitudinal and Hall conductivities}
Let us now turn to our experimental data.
We prepared a GaAs/AlGaAs 2DES sample with $\mu_m$ = 77 m$^2$/(Vs) and $n_e$ = 2.1$\times$10$^{15}$ m$^{-2}$, hence $\varepsilon_\mathrm{F}$ = 7.5 meV, shaped in a Hall bar geometry by photolithography. The quantum mobility $\mu_q$ = 7.1 m$^2$/(Vs) was determined from the damping of the amplitudes $\Delta \rho_\mathrm{SdH}$ of the Shubnikov-de Haas (SdH) oscillation at low magnetic fields, $\Delta \rho_\mathrm{SdH}(B)/\rho_0 = C \exp [-\pi/(\mu_qB)]$ with $\rho_0$ the resistivity at $B=0$ \cite{Coleridge91}. It was pointed out in Ref.\ \cite{Coleridge91} that the prefactor $C$ equals 4 in a homogeneous 2DES and the deviation from the value is attributable to the inhomogeneity. We have verified that the SdH amplitudes in our sample were described by the above equation with $C = 4$ reasonably well, confirming that inhomogeneity is minimal in our sample. (Note, however, that small inhomogeneity is inevitably present in a 2DES grown by molecular beam epitaxy, as will be discussed below.) Measurements were done in a dilution refrigerator equipped with a superconducting magnet at the base temperature ($\sim$15 mK), a temperature low enough for the approximation $k_\mathrm{B} T \ll \varepsilon_\mathrm{F}$ to be valid. The standard low-frequency (13 Hz) ac lock-in technique was employed for the resistivity measurement with a low excitation current (10 nA for $B < \sim 1$ T and 0.5 nA for higher magnetic fields) to prevent the electron heating. For the magnetic field sweep, we adopted very slow sweep rates (0.01 T/min for $B < \sim 1$ T and 0.1 T/min for higher magnetic fields), which, combined with a high data acquisition rate ($\sim$ 4 data points/s), allow us to acquire data points dense enough to perform the numerical differentiation with respect to $B$ reliably. The slow sweep rates are also favorable in avoiding the hysteresis in the superconducting magnet that obscures the exact value of the magnetic field felt by the sample. The longitudinal and the Hall resistances measured in our Hall bar sample are translated to resistivities $\rho_{xx}$ and $\rho_{xy}$ by using the geometrical factors of the Hall bar. Then we obtained $\sigma_{xx}$ and $\sigma_{xy}$ by numerically inverting the tensor, $\sigma_{xx} = \rho_{xx}/(\rho_{xx}^2+\rho_{xy}^2)$ and $\sigma_{xy} = \rho_{xy}/(\rho_{xx}^2+\rho_{xy}^2)$. As mentioned earlier, spin splitting can completely be neglected for $B \leq $ 0.5 T\@. Due to the spin degeneracy, the conductivities experimentally measured in this magnetic field range are simply twice as large as those without the spins; considering the spin degeneracy, the normalized conductivities are defined here as $\tilde{\sigma}_{\alpha \beta} = \sigma_{\alpha \beta}/(2 e^2/h)$.

%%%%%%% FIGURE 7 %%%%%%%%%
\begin{figure}[tb]
\begin{center}
\includegraphics[bbllx=50,bblly=25,bburx=710,bbury=1000,width=8.5cm]{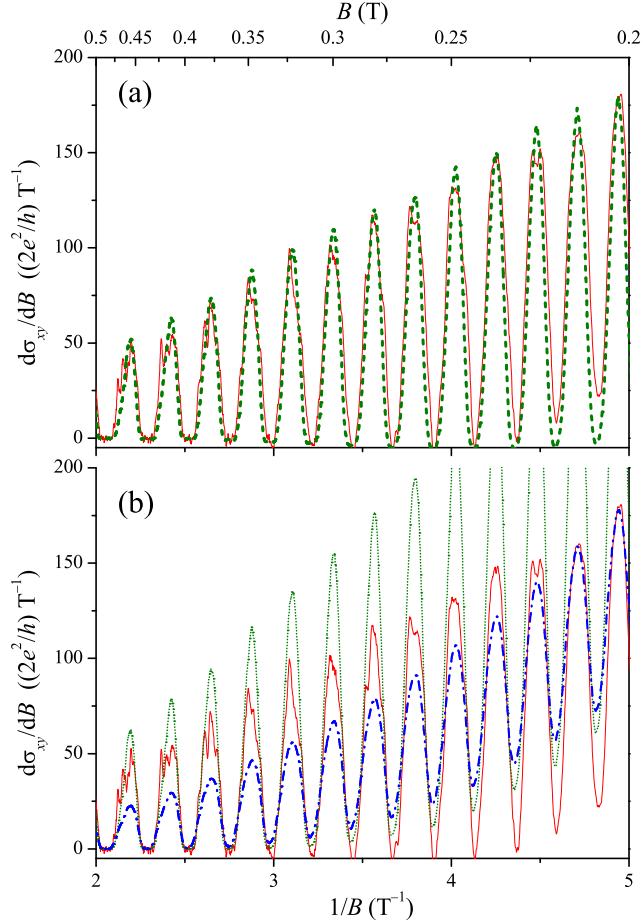}
\end{center}
\caption{\label{expfigs} Experimental traces to be compared with Fig.\ \ref{fundrelLF}: $\mathrm{d}\tilde{\sigma}_{xy}/\mathrm{d}B$ deduced by numerical differentiation of experimentally obtained $\sigma_{xy}$ [thin solid red line, plotted in both (a) and (b)], $\mathrm{d}\tilde{\sigma}_{xy}/\mathrm{d}B$ approximated by Eq.\ (\ref{fundrelLRn}) [thick dashed green line in (a)] or by Eq.\ (\ref{fundrelLR}) [thin dotted green line in (b)] calculated using experimentally obtained $\sigma_{xx}$. The r.h.s.\ of Eq.\ (\ref{chang2}) with the experimentally obtained $\sigma_{xx}$ and $\beta = 2 \tau_m / \tau_q$ is also plotted by dot-dashed blue line in (b). The traces are separately plotted in (a) and/or (b) for clarity.}
\end{figure}
%%%%%%%%%%%%%%%%%%%%%

In Fig.\ \ref{expfigs}, we show $\mathrm{d}\tilde{\sigma}_{xy}/\mathrm{d}B$ attained by the numerical differentiation of experimentally obtained $\sigma_{xy}$, and $\mathrm{d}\tilde{\sigma}_{xy}/\mathrm{d}B$ approximated by Eqs.\ (\ref{fundrelLR}) and (\ref{fundrelLRn}) using experimentally acquired $\sigma_{xx}$. It can be seen by comparing Figs.\ \ref{fundrelLF} and \ref{expfigs} that our theory reproduces the experimentally obtained traces remarkably well. Note that the same vertical scale is used for the two figures. Both figures reveal that the approximation by Eq.\ (\ref{fundrelLR}) progressively worsens with decreasing magnetic field, while Eq.\ (\ref{fundrelLRn}) remains a good approximation over the magnetic field range shown in the figure. We want to emphasize that the good quantitative agreement, demonstrated in Fig.\ \ref{expfigs}, between $\mathrm{d}\tilde{\sigma}_{xy}/\mathrm{d}B$ directly deduced from $\sigma_{xy}$ and that approximated by Eq.\ (\ref{fundrelLRn}) using $\sigma_{xx}$ is achieved without any fitting parameter. In Fig.\ \ref{expfigs}, we also plot the r.h.s.\ of Eq.\ (\ref{chang2}), a more conventional empirical relation. For the coefficient $\beta$, we adopted the relation $\beta =2 \tau_m / \tau_q = 2 \mu_m / \mu_q$ proposed by Coleridge \textit{et al.} \cite{Coleridge94}. We can see that Eq.\ (\ref{fundrelLRn}) describes the relation between $\sigma_{xx}$ and $\sigma_{xy}$ much better than Eq.\ (\ref{chang2}). It is clear from the figure that even if we use $\beta$ as a fitting parameter, agreement by Eq.\ (\ref{chang2}) cannot be improved very much.

%%%%%%% FIGURE 8 %%%%%%%%%
\begin{figure}[tb]
\begin{center}
\includegraphics[bbllx=50,bblly=25,bburx=710,bbury=1000,width=8.5cm]{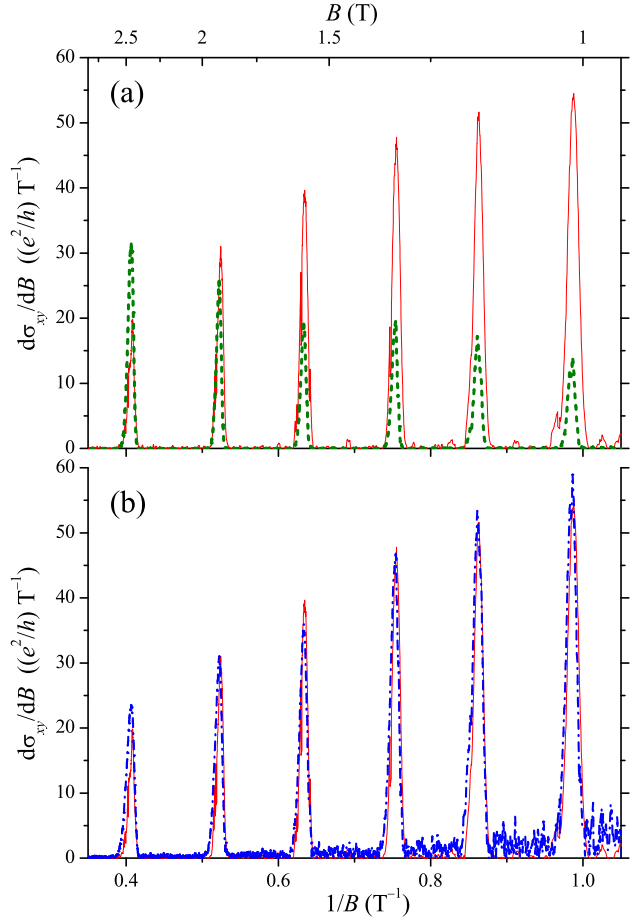}
\end{center}
\caption{\label{expfigsH} Experimental traces for higher magnetic-field range: $\mathrm{d}\tilde{\sigma}_{xy}/\mathrm{d}B$ deduced by numerical differentiation of experimentally obtained $\sigma_{xy}$ [thin solid red line, plotted in both (a) and (b)], $\mathrm{d}\tilde{\sigma}_{xy}/\mathrm{d}B$ approximated by Eq.\ (\ref{fundrelLR}) calculated using experimentally obtained $\sigma_{xx}$ [thick dashed green line in (a)], and the r.h.s.\ of Eq.\ (\ref{chang2}) with the experimentally obtained $\sigma_{xx}$ and $\beta \simeq$ 400 [dot-dashed blue line in (b)]. The traces are separately plotted in (a) and/or (b) for clarity.}
\end{figure}
%%%%%%%%%%%%%%%%%%%%%

For higher magnetic fields, spin splitting manifests itself as the splitting of the peaks in $\sigma_{xx}$ and $d \sigma_{xy} / dB$. The peaks take place at the conditions $\varepsilon_\mathrm{F} = E_N + g^* \sigma \mu_\mathrm{B} B$ [$N$ = 0, 1, 2,...,  $\sigma = \pm 1/2$, and $g^*$ represents the $g$-factor including ($B$-dependent) exchange enhancement], instead of $\varepsilon_\mathrm{F} = E_N$ in the spin-degenerate case, and therefore Eqs.\ (\ref{sgmxxLR}) and (\ref{sgmxyLR}) no longer describe the positions of peaks or steps between adjacent plateaus correctly. Nevertheless, concurrent occurrence of peaks in $\sigma_{xx}$ and in $d \sigma_{xy} / dB$ still allows us an attempt to see the applicability of Eq.\ (\ref{fundrelLR}), as shown in Fig.\ \ref{expfigsH}. Here  $\tilde{\sigma}_{\alpha \beta}=\sigma_{\alpha \beta}/(e^2/h)$ again since spin degeneracy is now lifted. We see that Eq.\ (\ref{fundrelLR}) reproduces roughly the right order of magnitude for the height of the peaks in $d \sigma_{xy} / dB$, although the increase in the peak height with increasing magnetic field for 1 T $\leq B \leq  $ 2.5 T is at obvious variance with the behavior of $d \sigma_{xy} / dB$. The discrepancy is mainly ascribable to the deviation of experimental peak heights in $\sigma_{xx}$ from the $\propto 1/B$ dependence inferred from Eq.\ (\ref{sgmxxLR}). By contrast, we find that our experimental result is well described by Eq.\ (\ref{chang2}) in accordance with previous studies \cite{Rotger89,Morawicz90,Stormer92,Coleridge94,Allerman95,Tieke97,Pan05}, albeit with the value of the parameter $\beta \simeq$ 400 roughly 20 times larger than $2 \tau_m/\tau_q$. 

\section{Discussion\label{Disc}}
The relation between $\sigma_{xx}$ and $\sigma_{xy}$ is already implicit in Eqs.\ (\ref{eq4}) and (\ref{eq5}), since both of the components derive from the same set of Green's function $G_N(\varepsilon)$ ($N$ = 0, 1, 2,...), or from the same DOS; note that once the DOS is given, both imaginary and real parts of $G_N(\varepsilon)$ are known by Eq.\ (\ref{eq2}) and by the Kramers-Kronig relation, respectively. We have shown in Sec.\ \ref{Relation} that the relation can be explicitly written down as Eq.\ (\ref{eq21}), or approximately as Eq.\ (\ref{eq22}), if we assume a simple form given by Eq.\ (\ref{eq4a}) for the Green's function corresponding to the Lorentzian broadening of the Landau levels Eq.\ (\ref{eq2a}). It might be argued that the expression Eq.\ (\ref{eq4a}) is too crude to represent a 2DES under magnetic field. We expect, however, that improvement in $G_N(\varepsilon)$ does not alter the relation Eq.\ (\ref{eq21}) to a large extent [if we keep ourselves within the framework of the approximate relation between the conductivity tensor and the Green's function represented by Eqs.\ (\ref{eq4}) and (\ref{eq5})], so long as the resultant DOS does not significantly deviate from the Lorentzian line shape. An important point we would like to stress is that the relation Eq.\ (\ref{eq21}) is inherent in the expressions of $\sigma_{xx}$ and $\sigma_{xy}$ and requires no external source, e.g., the inhomogeneity in the electron density.

In Eqs.\ (\ref{eq4}) and (\ref{eq5}), we have neglected a number of effects known to take place in a 2DES subjected to a magnetic field. These include the localization, the formation of the edge states (stripes of compressible states parallel to the edge of the sample interleaved with incompressible regions), and the electron-electron interaction. We have also neglected spins altogether as mentioned in Sec.\ \ref{CompExp}. Due to the localization in the tails of Landau level peaks, the width of the peaks in $\sigma_{xx}$ will become narrower than what is shown in Figs.\ \ref{Fig-2} and \ref{sgmxxsgmxytr} for the high magnetic field region where overlap between adjacent Landau level peaks can be neglected. The electron-electron interaction will engrave additional minima on the peaks of $\sigma_{xx}$ between adjacent integral quantum Hall states for $N<2$ Landau levels via the fractional quantum Hall effect \cite{Tsui82}, and also affect the height and shape of the peaks for higher Landau levels through forming the (probably incomplete) charge density wave states \cite{Koulakov96,Moessner96}. For long-range impurity potential, the peaks will be altered also by the network of the compressible and incompressible stripes formed around valleys or hills of the impurity potential \cite{Chklovskii93}. Strictly speaking, therefore, our theory applies only to the low-magnetic field region where these effects are negligibly small. This is exactly the region we have employed in the comparison with the experimental result in Fig.\ \ref{expfigs}. The excellent agreement between the theory and experiment attests to the correctness of our theory were it not for the additional effects neglected in the theory. The slight difference in the line shape between theoretical (Fig.\ \ref{fundrelLF}) and experimental (Fig.\ \ref{expfigs}) traces, with the theoretical trace showing asymmetry between sharp maxima and rather rounded minima, is attributable to the use of constant $\varepsilon_\mathrm{F}$ in the theory; in the experiment, $\varepsilon_\mathrm{F}$ is expected to oscillate with magnetic field to keep the electron density $n_e$ constant, resulting in more symmetric peaks and dips \cite{Endo08SdHHH}. 

In the higher magnetic-field regime, we envisage better agreement between theoretical and experimental results by modifying our theory to include the effects neglected in the present paper listed above, which is the subject of our future study. In the high-magnetic-field regime, however, we are unable to rule out the possibility that the inhomogeneity in $n_e$ is the dominant source of the experimentally observed relation Eq.\ (\ref{chang}) [or Eq.\ (\ref{chang2}) as shown in Fig.\ \ref{expfigsH}], as suggested by previous studies \cite{Simon94,Ilan06,Pan05}; the effect of the inhomogeneity is expected to gain more significance at higher magnetic fields, since the difference in the Hall resistivity $\Delta \rho_{xy}$ between two points differing in the electron density by $\Delta n_e$, $\Delta \rho_{xy} \simeq \Delta n_e B / (n_e^2 e)$, increases with $B$. Note that, in realistic samples, both microscopic inhomogeneity owing to the random distribution of the dopants and macroscopic inhomogeneity resulting from the technical difficulties in the molecular beam epitaxy are virtually impossible to be completely eliminated.

\section{Conclusions\label{Conc}}
We have calculated the diagonal ($\sigma_{xx}$) and off-diagonal ($\sigma_{xy}$) components of the conductivity tensor in the quantum Hall system by the linear response theory, neglecting the correction from the current vertex part. A Lorentzian line shape with the width $\Gamma$ independent of the magnetic field was assumed for the broadening of the Landau levels by the short-range impurity potential. The corresponding simple approximation for the Green's function Eq.\ (\ref{eq4a}) allowed us to obtain analytic formulas for both $\sigma_{xx}$ and $\sigma_{xy}$, given by Eqs.\ (\ref{eq20a}) and (\ref{eq20b}) respectively, for $k_\mathrm{B} T \ll \varepsilon_\mathrm{F}$. The formulas asymptotically approach the semiclassical formulas at low-magnetic fields. Inspection of the formulas reveals that $\mathrm{d} \sigma_{xy}/\mathrm{d}B$ is proportional to $B \sigma_{xx}^2$ [Eq.\ (\ref{eq22})] at high magnetic fields where $\Gamma \ll \hbar \omega_c$. This comprises a possible alternative route to explain, without resorting to the inhomogeneity in the electron density, the well-known empirical relation between $\sigma_{xx}$ and $\sigma_{xy}$.

To account for the long-range nature of the impurity potential in a GaAs/AlGaAs 2DES, slight modification was made by introducing two types of scattering times, the quantum scattering time $\tau_q$ and the momentum relaxation time $\tau_m$, yielding Eqs.\ (\ref{sgmxxLR}) and (\ref{sgmxyLR}) for $\sigma_{xx}$ and $\sigma_{xy}$, respectively. The resultant relation between the two components, Eq.\ (\ref{fundrelLRn}), is found to be in quantitative agreement with the experimental result obtained in the GaAs/AlGaAs 2DES at the magnetic field range where the spin splitting, the localization, the formation of the edge states, the electron-electron interaction, etc., can be neglected.

\ack
This work was supported by Grant-in-Aid for Scientific Research (B) (20340101) from the Ministry of Education, Culture, Sports, Science and Technology (MEXT) and by the National Institutes of Natural Sciences undertaking Forming Bases for Interdisciplinary and International Research through Cooperation Across Fields of Study and Collaborative Research Program (No.~NIFS08KEIN0091). Three of the authors (R.S., N.H. and H.N.) are grateful to generous supports from The Thermal and Electric Energy Technology Foundation, Research Foundation for Materials Science, and The Iketani Foundation, and to the support by Grant-in-Aid for Exploratory Research (17654073) from MEXT\@.  A. E. acknowledges the financial support by Grant-in-Aid for Scientific Research (C) (18540312) from MEXT.

\appendix

\section{Conductivity tensor in the weak magnetic field limit\label{WeakMag}}
In the derivation of Eqs.\ (\ref{eq20a}) and (\ref{eq20b}), we have made no assumption on the strength of the magnetic field. Therefore the equations should, in the low-field limit, asymptotically coincide with the well-known semiclassical expressions
\begin{eqnarray}
\sigma_{xx}^\mathrm{SC} & = & \frac{\sigma_0}{1+(\omega_c \tau)^2}  \\
\sigma_{xy}^\mathrm{SC} & = & -\frac{\sigma_0 \omega_c \tau}{1+(\omega_c \tau)^2}  = -\frac{n_e e}{B}+\frac{1}{\omega_c \tau} \sigma_{xx}^\mathrm{SC},
\end{eqnarray}
with $\sigma_0$ = $n_e e^2 \tau /m^*$ = $\varepsilon_\mathrm{F} e^2 \tau / (2 \pi \hbar^2)$ or, in the normalized forms,
\begin{eqnarray}
\tilde{\sigma}_{xx}^\mathrm{SC} & = & \frac{2 \gamma}{1+4\gamma^2} \left( X_\mathrm{F}+\frac{1}{2} \right)  \\
\tilde{\sigma}_{xy}^\mathrm{SC} & = & -\frac{1}{1+4\gamma^2} \left( X_\mathrm{F}+\frac{1}{2} \right)  = -\left( X_\mathrm{F}+\frac{1}{2} \right) + 2 \gamma \tilde{\sigma}_{xx}^\mathrm{SC}.
\end{eqnarray}
Since $\mathrm{Fsinh}(X_\mathrm{F},{\gamma}) \rightarrow 1$ with $\gamma \rightarrow \infty$, it is ready to see $\tilde{\sigma}_{xx} \rightarrow \tilde{\sigma}_{xx}^\mathrm{SC}$ with $B \rightarrow 0$ in Eq.\ (\ref{eq20a}).
From Eqs.\ (\ref{eq20a}) and (\ref{eq20b}), we find 
\begin{equation}
\label{eq18a}
\tilde{\sigma}_{xy} (X_\mathrm{F},{\gamma})  = - {\rm IFsinh}(X_\mathrm{F},\gamma) 
 - {\gamma} {\rm Fsin}(X_\mathrm{F},\gamma)  + 2 {\gamma} \tilde{\sigma}_{xx} (X_\mathrm{F},{\gamma}).
\end{equation} 
Noting that $\mathrm{IFsinh}(X_\mathrm{F},{\gamma}) \rightarrow (X_\mathrm{F}+1/2)$ and $\gamma \mathrm{Fsin}(X_\mathrm{F},{\gamma}) \rightarrow 0$ with $\gamma \rightarrow \infty$, we can also perceive $\tilde{\sigma}_{xy} \rightarrow \tilde{\sigma}_{xy}^\mathrm{SC}$ with $B \rightarrow 0$.

\bibliography{FundRel}

\end{document}